\documentclass[10pt,journal,compsoc]{IEEEtran}

\ifCLASSOPTIONcompsoc
\usepackage[nocompress]{cite}
\else
\usepackage{cite}
\fi

\usepackage[pdftex]{graphicx}
\graphicspath{{./}}
\DeclareGraphicsExtensions{.png}
\usepackage{caption}

\ifCLASSOPTIONcompsoc
\usepackage[caption=false,font=footnotesize,labelfont=sf,textfont=sf]{subfig}
\else
\usepackage[caption=false,font=footnotesize]{subfig}
\fi

\usepackage{subfig}
\usepackage{times}
\usepackage{amsthm}
\usepackage{amsfonts}
\usepackage{mathtools,amsmath}
\usepackage{url}
\usepackage{color, colortbl}
\usepackage{enumitem}

\DeclareMathOperator*{\argmin}{arg\,min}

\usepackage{pgfplots}
\usepgfplotslibrary{patchplots}
\pgfplotsset{compat=newest}
\usepackage{filecontents}
\usepackage{tikz}
\usetikzlibrary{spy}
\usepackage{pifont}
\usepackage{xcolor}
\definecolor{myblue}{RGB}{0,29,119}
\definecolor{graphs}{RGB}{242,235,234}
\definecolor{bblue}{HTML}{4F81BD}
\definecolor{rred}{HTML}{C0504D}
\definecolor{ggreen}{HTML}{9BBB59}
\definecolor{ppurple}{HTML}{9F4C7C}
\definecolor{orange}{HTML}{FF7F00}
\usepgfplotslibrary{units}
\usetikzlibrary{spy,backgrounds}
\usepackage{pgfplotstable}

\pgfplotstableread{
0.0    2.0
0.0   -3.5
}\datatable
\usepackage{multirow}
\usepackage{caption}

\newtheorem{theorem}{Theorem}
\newtheorem{prop}{Property}[section]
\newtheorem{defn}{Definition}[section]

\begin{document}

\title{Energy of Computing on Multicore CPUs: Predictive Models and Energy Conservation Law}

\author{Arsalan~Shahid,
	Muhammad~Fahad,
	Ravi~Reddy~Manumachu,
	and~Alexey~Lastovetsky,~\IEEEmembership{Member,~IEEE}
	\IEEEcompsocitemizethanks{\IEEEcompsocthanksitem A.Shahid, M. Fahad, R. Reddy and A. Lastovetsky are with the School of Computer Science, University College Dublin, Belfield, Dublin 4, Ireland.\protect\\
		E-mail: arsalan.shahid@ucdconnect.ie, muhammad.fahad@ucdconnect.ie, ravi.manumachu@ucd.ie, alexey.lastovetsky@ucd.ie}
	\thanks{}}

\markboth{Energy of Computing on Multicore CPUs: Predictive Models and Energy Conservation Law}
{Energy of Computing on Multicore CPUs: Predictive Models and Energy Conservation Law}

\IEEEtitleabstractindextext{%
\begin{abstract}
Energy is now a first-class design constraint along with performance in all computing settings. Energy predictive modelling based on performance monitoring counts (PMCs) is the leading method used for prediction of energy consumption during an application execution. We use a model-theoretic approach to formulate the assumed properties of existing models in a mathematical form. We extend the formalism by adding properties, heretofore unconsidered, that account for a limited form of energy conservation law. The extended formalism defines our theory of \emph{energy of computing}. By applying the basic practical implications of the theory, we improve the prediction accuracy of state-of-the-art energy models from 31\% to 18\%. We also demonstrate that use of state-of-the-art measurement tools for energy optimization may lead to significant losses of energy (ranging from 56\% to 65\% for applications used in experiments) since they do not take into account the \emph{energy conservation} properties.
\end{abstract}

}

\maketitle

\IEEEpeerreviewmaketitle

\IEEEraisesectionheading{\section{Introduction}\label{sec:introduction}}
Energy is now a first-class design constraint along with performance in all computing settings \cite{Barroso2007,DOE2010} and a serious environmental concern \cite{Smarr2010}. Accurate measurement of energy consumption during an application execution is key to application-level energy minimization techniques \cite{lastovetsky2016new}, \cite{Reddy2016TC}, \cite{manumachu2018parallel}, \cite{reddy2019design}. There are three popular approaches to providing it \cite{fahad2019comparative}: a). System-level physical measurements using external power meters, b). Measurements using on-chip power sensors, and c). Energy predictive models. The first approach lacks the ability to provide fine-grained component-level decomposition of the energy consumption of an application. This is essential to finding energy-efficient configuration of the application. The second approach is not accurate enough for the use in application-level energy optimization methods \cite{fahad2019comparative}.

Energy predictive modelling emerged as the pre-eminent alternative. The existing models predominantly use performance monitoring counts (PMCs) as predictor variables. PMCs are special-purpose registers provided in modern microprocessors to store the counts of software and hardware activities. A pervasive approach is to determine the energy consumption of a hardware component based on linear regression of the PMC counts in the component during an application run. The total energy consumption is then calculated as the sum of these individual consumptions.

In this work, we summarize and generalize the assumptions behind the existing work on PMC-based energy predictive modelling. We use a model-theoretic approach to formulate the assumed properties of the existing models in a mathematical form. We extend the formalism by adding properties, heretofore unconsidered, that are basic implications of the universal energy conservation law. The new properties are intuitive and have been experimentally validated. The extended formalism defines our theory of \emph{energy of computing}. Using the theory, we prove that an energy predictive model is linear if and only if its each PMC parameter is \emph{additive} in the sense that the PMC for a serial execution of two applications is the sum of PMCs for the individual execution of each application. 

Basic practical implications of the theory include an \emph{additivity} test identifying model parameters suitable for more reliable energy predictive modelling and constraints for models (For example: zero intercept and positive coefficients for linear regression models) that disallow violation of energy conservation properties. We incorporate these implications in the state-of-the-art models and study their prediction accuracy using a strict experimental methodology on a modern Intel multicore processor. 

As the first step, we test the \emph{additivity} of PMCs offered by the Likwid \cite{treibig2010likwid} package for compound applications. We show that all the PMCs fail the \emph{additivity} test where the input tolerance is 5\%. We observe that a PMC can be \emph{non-additive} with error as high as 3075\% and there are many PMCs where the error is over 100\%. This suggests that the use of highly \emph{non-additive} PMCs as predictor variables can impair the prediction accuracy of the models.

To understand the causes of the \emph{non-additivity}, we study the behaviour of PMCs with different numbers of threads/cores used in applications. We demonstrate a rise in the number of \emph{non-additive} PMCs with the increase in number of cores employed in the application. We consider this to be an inherent trait of a modern multicore computing platform because of its severe resource contention and non-uniform memory access (NUMA).

We select six PMCs which are common in the state-of-the-art models and which are highly correlated with dynamic energy consumption. All the PMCs fail the \emph{additivity} test for input tolerance of 5\%; one PMC is comparatively more \emph{additive} than the rest. We construct seven linear regression models, $\{A,B,...,G\}$. All the models have zero intercept and positive coefficients. They incorporate basic sanity checks that disallow violations of energy conservation property in our theory of \emph{energy of computing}.

ModelA employs all the selected PMCs as predictor variables. ModelB is based on five most \emph{additive} PMCs. ModelC uses four most \emph{additive} PMCs and so on until ModelF containing the highest \emph{additive} PMC. ModelG is based on three PMCs most correlated with dynamic energy consumption. We compare the prediction accuracies of these seven models plus \emph{Intel RAPL} (Running Average Power Limit) \cite{IntelRAPL} against the system-level physical measurements from power meters using \emph{HCLWattsUp}, which we consider to be the ground truth. We demonstrate that as we remove highly non-additive PMCs one by one from the models, their prediction accuracy improves. ModelE, which employs two most \emph{additive} PMCs has the best average prediction accuracy. Even though ModelF contains the highest additive PMC, it fares poorly due to poor linear fit thereby suggesting the perils of pure fitting exercise. \emph{RAPL}'s average prediction accuracy is equal to that of ModelA. ModelG fares better than \emph{RAPL} and ModelA.

Therefore, we conclude that use of highly \emph{additive} PMCs is crucial to good prediction accuracy of energy predictive models. Indeed, if PMCs used in the model are all non-additive with an error of 100\%, then the predictive error of the model cannot be less than 100\%.

Finally, to demonstrate the importance of the accuracy of energy measurements, we study optimization of a parallel matrix-matrix multiplication application for dynamic energy using two measurement methods. The first uses \emph{IntelRAPL} \cite{IntelRAPL} which is a popular mainstream tool. The second is based on system-level physical measurements using power meters (\emph{HCLWattsUp} \cite{HCLWattsUp}) which we believe are accurate. We show that using \emph{IntelRAPL} measurements instead of \emph{HCLWattsUp} ones will lead to significant energy losses ranging from 34\% to 67\% for matrix sizes used in the experiments.

The main original contributions of this work are:

\begin{itemize}
	\item Theory of energy of computing and its practical implications, which include an \emph{additivity} test for model parameters and constraints for model coefficients, that can be used to improve the prediction accuracy of energy models.
	\item Improvements to prediction accuracy of the state-of-the-art energy models using the practical implications of our theory of energy of computing.
	\item Study demonstrating significant energy losses incurred due to employment of inaccurate energy measuring tools (in energy optimization methods) since they do not take into account the \emph{energy conservation} properties.
\end{itemize}

We organize the rest of this paper as follows. We present terminology related to energy predictive models. This is followed by overview of our formal theory of \emph{energy of computing}. Then, we present experimental results followed by survey of related work and conclusion.

\section{Terminology}\label{sec:dynamicenergyrationale}

There are two types of power consumptions in a component: dynamic power and static power. Dynamic power consumption is caused by the switching activity in the component's circuits. Static power or idle power is the power consumed when the component is not active or doing work. From an application point of view, we define dynamic and static power consumption as the power consumption of the whole system with and without the given application execution. From the component point of view, we define dynamic and static power consumption of the component as the power consumption of the component with and without the given application utilizing the component during its execution.

There are two types of energy consumptions, static energy and dynamic energy. We define the static energy consumption as the energy consumption of the platform without the given application execution. Dynamic energy consumption is calculated by subtracting this static energy consumption from the total energy consumption of the platform during the given application execution. If $P_S$ is the static power consumption of the platform, $E_T$ is the total energy consumption of the platform during the execution of an application, which takes $T_E$ seconds, then the dynamic energy $E_D$ can be calculated as,

\begin{equation} \label{eq:DynamicEnergy}
\begin{aligned}
E_D = E_T - (P_S \times T_E)
\end{aligned}
\end{equation}

In this work, we consider only the dynamic energy consumption. We describe the rationale behind using dynamic energy consumption in the Appendix A.

\section{Energy Predictive Models of Computing: Intuition, Motivation, and Theory} \label{sec:energy-conservation}

We summarize and generalize the assumptions behind the current work on PMC-based power/energy modelling. We use a model-theoretic approach to formulate the assumed properties of these models in a mathematical form. Then we extend the formalism by adding properties, which are intuitive and which we have experimentally validated but have never been considered previously. The properties are manifestations of the fundamental physical law of energy conservation. We introduce two definitions based on the properties of the extended model, called \emph{weak composability} and \emph{strong composability}. An energy predictive model satisfying all the properties of the extended model is termed a \emph{consistent} energy model. The extended model and the two definitions define our theory of energy predictive models of computing.

Finally, we mathematically derive properties of \textit{linear} consistent energy predictive models. We prove that a consistent PMC-based energy model is linear if and only if it is strongly composable with each PMC variable being additive. The practical implication of this theoretical result is that each PMC variable of a linear energy predictive model must be \emph{additive}. The significance of this property is that it can be efficiently tested and hence used in practice to identify PMC variables that must not be included in the model. The notation and the terminology used in the proposed theory is given in Table \ref{table:nomenclature}.

\begin{table*}[!htbp]
	\centering
	\caption{Notation and terminology used in the theory of energy predictive models of computing.}
	\label{table:nomenclature}
	\begin{tabular}{|p{5cm}|p{10cm}|}
		\hline
		\textbf{Notation} & \textbf{Description} \\ \hline
		$A$, $B$, ... & Base applications \\ \hline
		$A \oplus B$ & Compound application of the base applications $A$ and $B$ \\ \hline
		$\mathcal{A}$ & Set of applications \\ \hline		
		$E(A)$ & Energy consumption of application $A$ \\ \hline
		$E(A \oplus B)$ & Energy consumption of compound application $A \oplus B$ \\ \hline
		$p=\{p_k\}_{k=1}^{n},q=\{q_k\}_{k=1}^{n} \in \mathbb R_{\ge 0}^n$ & PMC vectors $p$ and $q$ \\ \hline
		$\mathcal{NULL}=\{0\}_{k=1}^{n}$ & A null vector of PMCs \\ \hline		
		$f_{\!_E}: \mathbb R_{\ge 0}^n \rightarrow \mathbb R_{\ge 0}$ & A PMC-based energy predictive model \\ \hline
		$f_{\!_E}(a)$ & Energy value for the input PMC vector $a$ \\ \hline
		$\mathcal{O}$ & Set of binary operators \\ \hline
		$a_k \circ_{\!_{A}\!_{B}\!_{,k}} b_k$ & Binary operator $\circ_{\!_{A}\!_{B}\!_{,k}}$ combining the k-th PMCs $a_k$ and $b_k$ in the PMC vectors $a$ and $b$ for the applications $A, B \in \mathcal{A}$, respectively\\ \hline
		$\{\circ_{\!_{A}\!_{B}\!_{,1}},\cdots,\circ_{\!_{A}\!_{B}\!_{,n}}\}$ & Set of binary operators combining the PMC vectors for the applications $A, B \in \mathcal{A}$ \\ \hline		
	\end{tabular}
\end{table*}

\subsection{Intuition and Motivation}

The essence of PMC-based energy predictive models is that an application run can be accurately characterized by a $n$-vector of PMCs over $\mathbb R_{\ge 0}$. Any two application runs characterized by the same PMC vector are supposed to consume the same amount of energy. The applications in these runs may be different, but the same computing environment is always assumed. Thus, PMC-based models are computer system-specific. 

Based on these assumptions, any PMC-based energy model is formalized by a set of PMC vectors over $\mathbb R_{\ge 0}$, and a function, $f_{\!_E}: \mathbb R_{\ge 0}^n \rightarrow \mathbb R_{\ge 0}$, mapping these vectors in the set to energy values. \textit{No other properties of the set and the function are assumed.}

In this work, we extend this model by adding properties that characterize the serial execution of two applications. To aid the exposition, we follow some notation and terminology. A \emph{compound application} is defined as the serial execution of two applications, which we call the \emph{base} applications. If the base applications are $A$ and $B$, we denote their compound application by $A \oplus B$. We will refer solely to energy predictive models hereafter since there exists a linear functional mapping from PMC-based power predictive models to them. When we say energy consumption, we mean dynamic energy consumption. The energy consumption that is experimentally observed during the execution of an application $A$ is denoted by $E(A)$. The energy consumption of the compound application $A \oplus B$, $E(A \oplus B)$, is the energy consumption that is experimentally observed during the execution of the compound application.

First, we aim to reflect in the model the observation that in a stable and dedicated environment, where each run of the same application is characterized by the same PMC vector, for any two applications, the PMC vector of their serial execution will always be the same. To introduce this property, we add to the model a (infinite) set of applications denoted by $\mathcal{A}$. We postulate the existence of binary operators, $\mathcal{O}=\{\circ_{\!_{A}\!_{B}\!_{,k}}:\mathbb R_{\ge 0} \times \mathbb R_{\ge 0} \rightarrow \mathbb R_{\ge 0}, A,B \in \mathcal{A}, k \in [1,n]\}$ so that for each $A,B \in \mathcal{A}$ and their PMC vectors $a=\{a_k\}_{k=1}^{n},b=\{b_k\}_{k=1}^{n} \in \mathbb R_{\ge 0}^n$ respectively, the PMC vector of the compound application $A \oplus B$ will be equal to $\{a_k \circ_{\!_{A}\!_{B}\!_{,k}} b_k\}_{k=1}^{n}$.

Next, we introduce properties, which are manifestations of the universal energy conservation law. The following property essentially states that doing nothing (signified by a null vector of PMCs, $\mathcal{NULL}=\{0\}_{k=1}^{n} \in \mathbb R_{\ge 0}^n$) does not consume or generate energy,
\begin{align*}
f_{\!_E}(\mathcal{NULL}) = 0
\end{align*}

The following property postulates that an application with a PMC vector that is not $\mathcal{NULL}$ must consume some energy. The intuition behind this property is that since PMCs account for energy consuming activities of applications, an application with any energy consuming activity higher than zero activity (a $\mathcal{NULL}$ PMC vector), must consume more energy than zero.
\begin{align*}
\forall a \in \mathbb R_{\ge 0}^n \land a \neq \mathcal{NULL}, \text{  } f_{\!_E}(a) > 0
\end{align*}

Finally, we aim to reflect the observation that the consumed energy of compound application $A \oplus B$ is always equal to the sum of energies consumed by the individual applications A and B respectively,
\begin{equation}\label{energylaw}
E(A \oplus B) = E(A) + E(B)
\end{equation}
To introduce this property in the extended model, we postulate the following, 
\begin{align*}
\forall A,B \in \mathcal{A}, & \text{  } a=\{a_k\}_{k=1}^{n}, b=\{b_k\}_{k=1}^{n} \in \mathbb R_{\ge 0}^n, \text{  } \circ_{\!_{A}\!_{B}\!_{,k}} \in \mathcal{O}, \\
& f_{\!_E}(\{a_k \circ_{\!_{A}\!_{B}\!_{,k}} b_k\}_{k=1}^{n}) = f_{\!_E}(a) + f_{\!_E}(b)
\end{align*}

To summarize, while existing models are focused on abstract application runs and lack any notion of applications, we introduce this notion in the extended model. The additional structure introduced in the extended model allows one to prove the mathematical properties of energy predictive models.

\subsection{Formal Summary of Properties of Extended Model}

The formal summary of the properties of the extended model follows:

\begin{prop}[Inherited from Basic Model] \label{basicprop}
	
	An abstract application run is accurately characterized by a set of $n$-vector of PMCs over $\mathbb R_{\ge 0}$. A null vector of PMCs is represented by $\mathcal{NULL}=\{0\}_{k=1}^{n}$. There exists a function, $f_{\!_E}: \mathbb R_{\ge 0}^n \rightarrow \mathbb R_{\ge 0}$, mapping the vectors to energy values and $\forall p,q \in \mathbb R_{\ge 0}^n, p = q \implies f_{\!_E}(p)=f_{\!_E}(q)$.
	
\end{prop}

\begin{prop}[Weak Composability, Applications and Operators] \label{weakcomposabilityprop}
	
	There exists an application space, $(\mathcal{A}, \oplus)$, where $\mathcal{A}$ is a (infinite) set of applications and $\oplus$ is a binary function on $\mathcal{A}$, $\oplus: \mathcal{A} \times \mathcal{A} \rightarrow \mathcal{A}$. There exists a (infinite) set of binary operators, $\mathcal{O}=\{\circ_{\!_{P}\!_{Q}\!_{,k}}: \mathbb R_{\ge 0} \times \mathbb R_{\ge 0} \rightarrow \mathbb R_{\ge 0}, P,Q \in \mathcal{A}, k \in [1,n]\}$ so that for each $P,Q \in \mathcal{A}$ and their PMC vectors $p=\{p_k\}_{k=1}^{n},q=\{q_k\}_{k=1}^{n} \in \mathbb R_{\ge 0}^n$ respectively, the PMC vector of the compound application $P \oplus Q$ will be equal to $\{p_k \circ_{\!_{P}\!_{Q}\!_{,k}} q_k\}_{k=1}^{n}$.
	
\end{prop}

\begin{prop}[Zero Energy, Energy Conservation] \label{energyconservationprop1}
	
	$f_{\!_E}(\mathcal{NULL}) = 0$.
	
\end{prop}

\begin{prop}[Positive-definiteness, Energy Conservation] \label{energyconservationprop2}
	
	$\forall p \in \mathbb R_{\ge 0}^n \land p \neq \mathcal{NULL}, \text{  } f_{\!_E}(p) > 0$.
	
\end{prop}

\begin{prop}[Weak Composability, Energy Conservation] \label{energyconservationprop3}
	
	$\forall P,Q \in \mathcal{A}, \text{  } p=\{p_k\}_{k=1}^{n}, q=\{q_k\}_{k=1}^{n} \in \mathbb R_{\ge 0}^n, \text{  } \circ_{\!_{P}\!_{Q}\!_{,k}} \in \mathcal{O}, f_{\!_E}(\{p_k \circ_{\!_{P}\!_{Q}\!_{,k}} q_k\}_{k=1}^{n}) = f_{\!_E}(p) + f_{\!_E}(q)$.
	
\end{prop}

We term an energy predictive model satisfying all the above properties of the extended model a \emph{consistent} energy model.

\subsection{Strong Composability: Definition}

The definition of \emph{strong composability} of models follows:

\begin{defn} [Strong Composability] \label{strongcomposability}
	A consistent energy model is \emph{strongly composable} if $\forall P,Q,R,S \in \mathcal{A}, p=\{p_k\}_{k=1}^{n},q=\{q_k\}_{k=1}^{n},r=\{r_k\}_{k=1}^{n},s=\{s_k\}_{k=1}^{n} \in \mathbb R_{\ge 0}^n, k \in [1,n], \circ_{\!_{P}\!_{Q}\!_{,k}} = \circ_{\!_{R}\!_{S}\!_{,k}}$.
\end{defn}

The \emph{strong composability} property of a model essentially states that binary operators used in the model to compute PMC vectors of compound applications are not application specific.  In other words, the set $\mathcal{O}$ consists of only $n$ binary operators, one for each PMC parameter, $\mathcal{O} = \{ \circ_{k} \}_{k=1}^{n}$, so that for any $P,Q \in \mathcal{A}$ and their PMC vectors $p=\{p_k\}_{k=1}^{n},q=\{q_k\}_{k=1}^{n} \in \mathbb R_{\ge 0}^n$, the PMC vector of the compound application $P \oplus Q$ will be equal to $\{p_k \circ_{k} q_k\}_{k=1}^{n}$.

\subsection{Mathematical Analysis of Linear Energy Predictive Models Based on The Theory of Energy of Computing}

In this section, we mathematically derive properties of \textit{linear} consistent energy predictive models, that is, linear energy models satisfying properties (\ref{basicprop} to \ref{energyconservationprop3}). 

By definition, a model is \textit{linear} iff $f_{\!_E}(x)$ is a linear function.

To the best of our knowledge, all the state-of-the-art energy predictive models for multicore CPUs are based on linear regression. While they model total energy consumption, we consider dynamic energy consumption for reasons described in the Appendix \ref{denergy-rationale}. The mathematical form of these models can be stated as follows:
$\forall p=(p_k)_{k=1}^n, p_k \in \mathbb R_{\ge 0}$,
\begin{equation}\label{linear_energy_model}
f_{\!_E}(p) = \beta_0 + \beta \times p = \beta_0 + \sum_{k=1}^{n}\beta_k \times p_k
\end{equation}
where $\beta_0$ is called the model intercept, the $\beta=\{\beta_1,...,\beta_n\}$ is the vector of regression coefficients or the model parameters. In real life, there usually is stochastic noise (measurement errors). Therefore, the measured energy is typically expressed as 
\begin{equation}\label{linear_energy_model_reallife}
\tilde{f}_{\!_E}(p) = f_{\!_E}(p)  + \epsilon
\end{equation}
where the error term or noise $\epsilon$ is a Gaussian random variable with expectation zero and variance $\sigma^2$, written $\epsilon \sim \mathcal{N}(0,\sigma^{2})$. We will ignore the noise term in our mathematical proofs to follow.

\begin{theorem}
	If a linear energy predictive model (\ref{linear_energy_model}) is consistent, the model intercept must be zero and the model coefficients must be positive.
\end{theorem}
\begin{proof}
	From the energy conservation property \ref{energyconservationprop1}, 
	\begin{align*}
	\mathcal{NULL}=\{0\}_{k=1}^{n} \in \mathbb R_{\ge 0}^n, \text{ } & f_{\!_E}(\mathcal{NULL}) = 0 \\
	\implies & \beta_0 + \sum_{k=1}^{n}\beta_k \times 0 = 0 \\	
	\implies & \beta_0 = 0
	\end{align*}
	
	From the energy conservation property \ref{energyconservationprop2}, 
	\begin{align*}
	\forall k \in [1, n],  p &= \{0,...,0,p_k,0,...,0\} \land p \neq \mathcal{NULL}, \\
	& f_{\!_E}(p) > 0 \\
	\implies & \sum_{i=1}^{n}\beta_i \times p_i > 0 \\	
	\implies & \beta_k \times p_k > 0 \\
	\implies & \beta_k > 0 \text{ since } p_k > 0	
	\end{align*}	
\end{proof}

To summarize, a linear energy predictive model satisfying energy conservation properties (\ref{energyconservationprop1} and \ref{energyconservationprop2}) has a zero model intercept and positive model coefficients. Also as we only consider models satisfying property \ref{energyconservationprop1}, then the linearity of function $f_{\!_E}(x)$ can be equivalently defined as follows: for any $\alpha \in \mathbb R_{\ge 0}$ and $p, q \in \mathbb R_{\ge 0}^{n}$

\begin{equation} \label{eqn:linear1}
f_{\!_E}(p + q) = f_{\!_E}(p) + f_{\!_E}(q)
\end{equation}

and 

\begin{equation} \label{eqn:linear2}
f_{\!_E}(\alpha \times p) = \alpha \times f_{\!_E}(p)
\end{equation}

\begin{theorem} \label{ifproof}
	If a consistent energy model is linear, then it is strongly composable with O = \{+\}.
\end{theorem}
\begin{proof}
	From properties  \ref{weakcomposabilityprop} and \ref{energyconservationprop3} of \emph{weak composability}, we have
	\begin{align*}
	\forall P,Q \in \mathcal{A}, \forall k \in [1, n], p &= \{0,...,0,p_k,0,...,0\}, \\
	q=\{0,...,0,q_k,0,...,0\}: \\
	f_{\!_E}(\{0,...,0,p_k \circ_{\!_{P}\!_{Q}\!_{,k}} q_k,0,...,0\}) &= f_{\!_E}(p) + f_{\!_E}(q)
	\end{align*}
	
	Using the property (\ref{eqn:linear1}) of a linear predictive model,
	\begin{align*}
	& f_E(p + q) = f_E(p) + f_E(q) \\
	& \implies f_E(p + q) = f_{\!_E}(\{0,...,0, p_k \circ_{\!_{P}\!_{Q}\!_{,k}} q_k, 0,...,0\}) \\
	& \implies  f_E(\{0,..., p_k + q_k, 0,...,0\}) \\
	& \quad \quad = f_{\!_E}(\{0,...,0, p_k \circ_{\!_{P}\!_{Q}\!_{,k}} q_k, 0,...,0\}) \\
	& \implies p_k + q_k = p_k \circ_{\!_{P}\!_{Q}\!_{,k}} q_k \quad (\textrm{from linearity of} f_{E}(x)) \\
	& \implies \circ_{\!_{P}\!_{Q}\!_{,k}} = +	
	\end{align*}
	
	Therefore, if a consistent energy model is linear, then it is strongly composable with $O = \{+\}$.
\end{proof}

\begin{theorem} \label{converseproof}
	If a consistent energy model is strongly composable with O = \{+\} and function $f_{\!_E}(x)$ is continuous, then it is linear.
\end{theorem}
\begin{proof}
	First, we prove the first defining linearity property (\ref{eqn:linear1}),
	\begin{align*}    
	f_{\!_E}(p + q) = f_{\!_E}(p) + f_{\!_E}(q)
	\end{align*}
	for any $p,q \in \mathbb R_{\ge 0}^n$.
	
	As the model is \textit{strongly composable} with $O = \{ + \}$, then  
	\begin{align*}
	\forall P,Q \in \mathcal{A}, \forall k \in [1, n]:  \circ_{\!_{P}\!_{Q}\!_{,k}} = +
	\end{align*} 
	
	From property \ref{energyconservationprop3} of \emph{weak composability},
	\begin{align*}
	f_{\!_E}(\{p_k \circ_{\!_{P}\!_{Q}\!_{,k}} q_k\}_{k=1}^{n}) &= f_{\!_E}(p) + f_{\!_E}(q) \\
	\implies f_{\!_E}(p) + f_{\!_E}(q) &= f_{\!_E}(\{p_k \circ_{\!_{P}\!_{Q}\!_{,k}} q_k\}_{k=1}^{n}) \\
	\implies f_{\!_E}(p) + f_{\!_E}(q) &= f_{\!_E}(\{p_k + q_k\}_{k=1}^{n}) = f_E(p + q)
	\end{align*}
	
	This proves the first property of linearity. 
	
	We now prove the second defining property of linearity (\ref{eqn:linear2}),
	\begin{align*}    
	f_{\!_E}(\alpha \times p) = \alpha \times f_{\!_E}(p)
	\end{align*}
	for any $p \in \mathbb R_{\ge 0}^n$ and $\alpha \in \mathbb R_{\ge 0}$.	
	
	For any integer $m > 0$,
	\begin{align*}    
	f_{E}(m \times p) &= f_{E}(p + p + ... + p) \\
	&= f_{E}(p) + f_{E}(p) +...+ f_{E}(p) \\
	&= m \times f_{E}(p)
	\end{align*}
	
	For any integer $n > 0$,
	\begin{align*}    
	f_{E}(p) &= f_{E}(\frac{p}{n}) + f_{E}(\frac{p}{n}) + ... + f_{E}(\frac{p}{n}) \\
	&= n \times f_{E}(\frac{p}{n}) \\
	\implies \frac{1}{n} f_{E}(p) &= f_{E}(\frac{q}{n}) \\
	\end{align*}	
	
	Thus, for any rational $\frac{m}{n} > 0$, 
	\begin{align*}    
	\frac{m}{n}f_{E}(q) &= \frac{1}{n} f_{E}(q) + \frac{1}{n} f_{E}(q) + ... + \frac{1}{n} f_{E}(q) \\
	&= f_{E}(\frac{q}{n}) + f_{E}(\frac{q}{n}) + ... + f_{E}(\frac{q}{n} ) \\
	&= f_{E}(m \times \frac{q}{n}) = f_{E}(\frac{m}{n}q)
	\end{align*}
	
	By definition, any real number $\alpha$ is a limit of an infinite sequence of rational numbers. Consider a sequence $\{\alpha_k\}$ of positive rational numbers such that $\lim_{k \to +\infty} \alpha_k = \alpha$. Then,
	\begin{align*}    
	f_{E}(\alpha \times p) &= f_{E}((\lim_{k \to +\infty} \alpha_k) \times p) \\
	&= f_{E}(\lim_{k \to +\infty} (\alpha_k \times p)) \\
	&= \lim_{k \to +\infty} f_{E}(\alpha_k \times p) \text{ } (\textrm{from continuity of} f_{E}(x))
	\end{align*}
	
	As $\alpha_k$ are positive rational numbers, $f_{E}(\alpha_k \times p) = \alpha_k \times f_{E}(p)$. Therefore,
	\begin{align*}
	f_{E}(\alpha \times p) &= \lim_{k \to +\infty} (\alpha_k \times f_{E}(p)) \\
	&= f_{E}(p) \times \lim_{k \to +\infty} \alpha_k \\
	&= f_{E}(p) \times \alpha
	\end{align*}
\end{proof}

Therefore, we prove using theorem \ref{ifproof} and theorem \ref{converseproof} that a consistent energy model is linear if and only if it is strongly composable with $O = {+}$. A consistent PMC-based energy model is linear if and only if it is strongly composable, with each PMC variable being additive. The practical implication of this theoretical result is that each PMC variable of a linear energy predictive model must be additive.

\section{Experimental Results}

This section is divided into two parts. 

In the first part, we study the \emph{additivity} of PMCs for compound applications using an \emph{additivity} test. We analyse the impact on prediction accuracy of models using \emph{additive} and \emph{non-additive} PMCs as predictor variables.

In the second part, we study optimization of a parallel matrix-matrix application for dynamic energy using two measurement tools, \emph{IntelRAPL} \cite{IntelRAPL} which is a popular mainstream tool and system-level physical measurements using power meters (\emph{HCLWattsUp} \cite{HCLWattsUp}).

\subsection{Study of Additivity of PMCs}

\begin{figure*}
	\centering
	\includegraphics[width=6.5in]{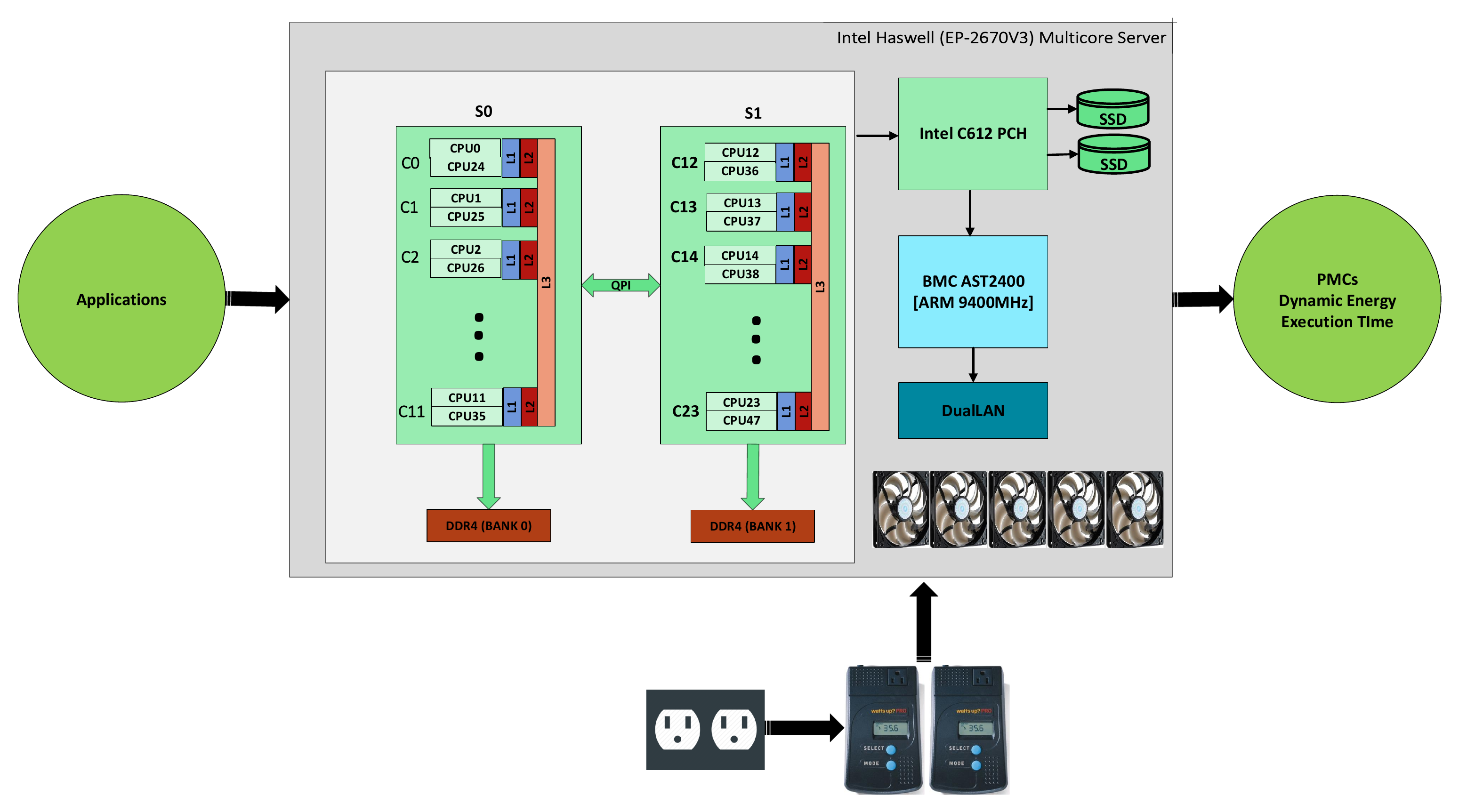}
	\caption{Experimental workflow to determine the PMCs on the Intel Haswell server.}
	\label{fig:work}
\end{figure*}

Our experimental platform is a modern Intel Haswell multicore server CPU whose specifications are given in the Table \ref{table:intel-haswell-multicore-CPU}. The experimental setup is illustrated in Figure~\ref{fig:work}. Our experimental testsuite (Table \ref{table:applications}) comprises of highly optimized applications (DGEMM, FFT) from Intel math kernel library (MKL), NAS parallel benchmarking suite (NPB), HPCG, and unoptimized matrix-matrix and matrix-vector multiplication applications.

\begin{table}[!t]
\footnotesize
	\caption{Specification of the Intel Haswell multicore CPU}
	\label{table:intel-haswell-multicore-CPU}
	\centering
	\begin{tabular}{|l|l|}
		\hline 
		\textbf{Technical Specifications} & \textbf{Intel Haswell Server} \\ \hline
		Processor & Intel E5-2670 v3 @2.30GHz \\ \hline
		OS & CentOS 7 \\ \hline
		Micro-architecture & Haswell \\ \hline
		Thread(s) per core & 2 \\ \hline	
		Cores per socket & 12 \\ \hline
		Socket(s) & 2 \\ \hline
		NUMA node(s) & 2 \\ \hline
		L1d cache & 32 KB \\ \hline
		L11 cache & 32 KB \\ \hline
		L2 cache & 256 KB \\ \hline
		L3 cache & 30720 KB \\ \hline
		Main memory & 64 GB DDR4 \\ \hline
		Memory bandwidth & 68 GB/sec \\ \hline
		TDP & 240 W \\ \hline	
		Idle Power & 58 W \\ \hline
	\end{tabular}
\end{table}

\begin{table}[!t]
	\centering
	\footnotesize
	\caption{List of Applications}
	\label{table:applications}
	\begin{tabular}{|p{1.5cm}|p{5cm}|}
		\hline
		\textbf{Application} & \textbf{Description} \\ \hline
		MKL FFT & Fast Fourier Transform \\ \hline
		MKL DGEMM & Dense Matrix Multiplication \\ \hline
		HPCG & High performance conjugate gradient \\ \hline
		NPB IS                & Integer Sort, Kernel for random memory access \\ \hline
		NPB LU                & Lower-Upper Gauss-Seidel solver \\ \hline
		NPB EP                & Embarrassingly Parallel, Kernel\\ \hline
		NPB BT                & Block Tri-diagonal solver \\ \hline
		NPB MG                & Multi-Grid on a sequence of meshes \\ \hline
		NPB FT                & Discrete 3D fast Fourier Transform \\ \hline
		NPB DC                & Data Cube \\ \hline
		NPB UA                & Unstructured Adaptive mesh, dynamic and irregular memory access, \\ \hline
		NPB CG                & Conjugate Gradient \\ \hline
		NPB SP                & Scalar Penta-diagonal solver \\ \hline
		NPB DT                & Data traffic \\ \hline
		\emph{stress}         & CPU, disk and I/O stress \\ \hline
		\emph{Naive MM}       & Naive Matrix-matrix multiplication \\ \hline
		\emph{Naive MV}       & Naive Matrix-vector multiplication \\ \hline
	\end{tabular}
\end{table}

For each application run, we measure the following: 1). Dynamic energy consumption, 2). Execution time, and 3). PMCs. The dynamic energy consumption during the application execution is measured using a \emph{WattsUp Pro} power meter and obtained programmatically via the HCLWattsUp interface \cite{HCLWattsUp}. The power meter is periodically calibrated using an ANSI C12.20 revenue-grade power meter, Yokogawa WT210.

We use \emph{Likwid} \cite{treibig2010likwid},\cite{LikwidHaswell} to obtain the PMCs. It offers 164 PMCs on our platform. We eliminate PMCs with counts less than or equal to 10. The eliminated PMCs have no significance on modelling energy consumption of our platform. The reduced set contains 151 PMCs. Collecting all these PMCs takes lots of time since only a limited number of PMCs can be obtained in a single application run due to the limited number of hardware registers dedicated to storing them. Therefore, each application must be executed about 53 times to collect all the PMCs.

\subsubsection{Steps to Ensure Reliable Experiments} \label{steps-reliable-experiments}

To ensure the reliability of our results, we follow a statistical methodology where a sample mean for a response variable is obtained from multiple experimental runs. The sample mean is calculated by executing the application repeatedly until it lies in the 95\% confidence interval and a precision of 0.025 (2.5\%) has been achieved. For this purpose, Student's t-test is used assuming that the individual observations are independent and their population follows the normal distribution. We verify the validity of these assumptions by plotting the distributions of observations.

The server is fully dedicated for the experiments. To ensure reliable energy measurements, we took following precautions: 

\begin{enumerate}
	\item \emph{HCLWattsUp} API \cite{HCLWattsUp} gives the total energy consumption of the server during the execution of an application using system-level physical measurements from the external power meters. This includes the contribution from components such as NIC, SSDs, fans, etc. To ensure that the value of dynamic energy consumption is purely due to CPUs and DRAM, we verify that all the components other than CPUs and DRAM are idle using the following steps:
	\begin{itemize}
		\item Monitoring the disk consumption before and during the application run. We ensure that there is no I/O performed by the application using tools such as \emph{sar}, \emph{iotop}, etc.
		\item Ensuring that the problem size used in the execution of an application does not exceed the main memory, and that swapping (paging) does not occur.
		\item Ensuring that network is not used by the application using monitoring tools such as \emph{sar}, \emph{atop}, etc.
		\item Bind an application during its execution to resources using cores-pinning and memory-pinning.
	\end{itemize}
	\item Our platform supports three modes to set the fans speed: \emph{minimum}, \emph{optimal}, and \emph{full}. We set the speed of all the fans to \emph{optimal} during the execution of our experiments. We make sure there is no contribution to the dynamic energy consumption from fans during an application run, by following the steps below:
	\begin{itemize}
		\item We continuously monitor the temperature of server and the speed of fans, both when the server is idle, and during the application run. We obtain this information by using Intelligent Platform Management Interface (IPMI) sensors.
		\item We observed that both the temperature of server and the speeds of the fans remained the same whether the given application is running or not.
		\item We set the fans at \emph{full} speed before starting the application run. The results from this experiment were the same as when the fans were run at \emph{optimal} speed.
		\item To make sure that pipelining, cache effects, etc, do not happen, the experiments are not executed in a loop and sufficient time (120 seconds) is allowed to elapse between successive runs. This time is based on observations of the times taken for the memory utilization to revert to base utilization and processor (core) frequencies to come back to the base frequencies.
	\end{itemize}
\end{enumerate}

\subsubsection{Ranking PMCs Using \emph{Additivity} Test}

We study the \emph{additivity} of PMCs offered by \emph{Likwid} using a test consisting of two stages. In the first stage, we determine if the PMC is deterministic and reproducible. In the second stage, we check if the PMC of compound application is equal to the sum of the values of corresponding PMC of base applications. A PMC must pass both stages to be called \emph{additive} for a given compound application on a given platform. 

First, we collect the values of the PMCs for the base applications by executing them separately. Next, we execute the \emph{compound} application and obtain its value of the PMC. If the PMC of the \emph{compound} application is equal to the sum of the PMCs of the base applications (with a tolerance of 5.0\%), we classify the PMC as potentially \emph{additive}. Otherwise, it is \emph{non-additive}.

For the experimental results, we prepare a dataset consisting of 60 compound applications composed from the base applications presented in Table \ref{table:applications}. No PMC is found to be additive within specified tolerance of 5\%. If we increase the tolerance to 20\%, 50 PMCs become \emph{additive}. Increasing the tolerance to 30\% makes 109 PMCs \emph{additive}. We observe that a PMC can be \emph{non-additive} with an error as high as 3075\% and there are many PMCs where the error is over 100\%.

Therefore, we conclude that all the PMCs fail the \emph{additivity} test with specified tolerance of 5\% on current multicore platforms.

\subsubsection{Evolution of Additivity of PMCs from Single-core to Multicore Architectures}

To identify the cause of this \emph{non-additivity}, we perform an experimental study to observe the \emph{additivity} of PMCs with different configurations of threads/cores employed in an application.

We choose for this study three applications: 1). MKL DGEMM, 2). MKL FFT and 3). \emph{naive} matrix-vector (MV) multiplication. We perform \emph{additivity} test for the applications for four different core configurations (2-core, 8-core, 16-core and 24-core). In the 2-core configuration, the application is pinned to one core of each socket. In the 8-core configuration, the application is pinned to four cores of each socket and so on. We design multiple compound applications from the chosen set of problem sizes. For each application and core configuration, we note the maximum percentage error for each PMC and count the number of \emph{non-additive} PMCs that exceed the input tolerance of 5\%.

Figure \ref{fig:pmcswithcores} shows the increase in \emph{non-additivity} of PMCs as the number of cores is increased for DGEMM, FFT and \emph{naive} MV. For DGEMM, 51 PMCs are \emph{non-additive} for 2-core configuration. The number increases to 126 for 24-core configuration. For FFT, the number increases from 61 to 146 and for \emph{naive} MV, the number increases from 22 to 58 from 2-core to 24-core configurations. The minimum number of \emph{non-additive} PMCs is for the 2-core configuration for each application.

Therefore, we conclude that the number of \emph{non-additive} PMCs increases with the increase in cores employed in an application execution because of severe resource sharing and contention.

\begin{table*}
\begin{tabular}{cc}
\centering
\pgfplotsset{width=7cm, height=5cm}
\begin{tikzpicture}
\begin{axis}[
   legend pos=south east,
   bar width=4mm, 
   enlargelimits=0.20,
   tick label style={font=\footnotesize},
   xticklabels={2,8,16,24},xtick=data,
   xlabel style={text width=3cm}, 
   xlabel=Number of Cores,
    ylabel=\emph{Non-Additive} PMCs,
]
\addplot+[sharp plot, red!80!white, sharp plot,mark=square*, mark options={scale=1}, line width=0.7mm] table [
    x=cores,
    y=pmcs
    ] {dgemm.dat};
\addplot+[ybar, fill=black!90!white, red!40!black] table [
    x=cores,
    y=pmcs
    ] {dgemm.dat};
   
\end{axis}
\end{tikzpicture}
&
\pgfplotsset{width=7cm, height=5cm}
\begin{tikzpicture}
\begin{axis}[
   xtick=data,
   legend pos=south east,
   bar width=4mm, 
   enlargelimits=0.20,
   xticklabels={2,8,16,24},xtick=data,
   tick label style={font=\footnotesize},
   xlabel style={text width=3cm}, 
   xlabel=Number of Cores,
    ylabel=\emph{Non-Additive} PMCs,
]
\addplot+[red!80!white, sharp plot,mark=square*, mark options={scale=1}, line width=0.7mm] table [
    x=cores,
    y=pmcs
    ] {fft.dat};
\addplot+[ybar, fill=black!90!white, red!40!black] table [
    x=cores,
    y=pmcs
    ] {fft.dat};
   
\end{axis}
\end{tikzpicture} \\
\textbf{A} & \textbf{B} \\
\multicolumn{2}{c}{
\pgfplotsset{width=7cm, height=5cm}
\begin{tikzpicture}
\begin{axis}[
   legend pos=south east,
   bar width=4mm, 
   enlargelimits=0.20,
   xticklabels={2,8,16,24},xtick=data,
   tick label style={font=\footnotesize},
   xlabel style={text width=3cm}, xlabel=Number of Cores,
    ylabel=\emph{Non-Additive} PMCs,
]
\addplot+[red!80!white, sharp plot,mark=square*, mark options={scale=1}, line width=0.7mm] table [
    x=cores,
    y=pmcs
    ] {mv.dat};
\addplot+[ybar, fill=black!90!white, red!40!black] table [
    x=cores,
    y=pmcs
    ] {mv.dat};
   
\end{axis}
\end{tikzpicture}}
\\
\multicolumn{2}{c}{\textbf{C}} \\
\end{tabular}
\captionof{figure} {Increase in number of \emph{non-additive} PMCs with threads/cores used in an application. \textbf{(A)}, \textbf{(B)}, and \textbf{(C)} shows \emph{non-additive} PMCs for Intel MKL DGEMM, Intel MKL FFT and \emph{naive} matrix-vector multiplication.}
\label{fig:pmcswithcores}
\end{table*}

\subsubsection{Improving Prediction Accuracy of Energy Predictive Models}

We select six PMCs common to the state-of-the-art models \cite{dolz2016analytical,haj2016fine,wang2014software,IcsiMartonosi2003,Li2003,Singh2009}. The PMCs ($\{X_1,\cdots,X_6\}$) are listed in the Table \ref{tab:pmcs-correlation}. They count floating-point and memory instructions and are considered to have a high positive correlation with energy consumption. They fail the \emph{additivity} test for an input tolerance of 5\%. $X_6$ is highly \emph{additive} compared to the rest. 

\begin{table*}[]
\centering
\caption{Correlation of PMCs with dynamic energy consumption ($E_{D}$). \textbf{(A)} List of selected PMCs for modelling with their \emph{additivity} test errors (\%). \textbf{(B)} Correlation matrix showing positive correlations of dynamic energy with PMCs. 100\% correlation is denoted by 1. $X_{4}$,$X_{5}$, and $X_{6}$ are highly correlated with $E_{D}$.}
\scriptsize
\label{tab:pmcs-correlation}
\begin{tabular}{c|c}
\begin{tabular}{p{4.6cm} p{1.7cm}}
\textbf{Selected PMCs}  &  \textbf{\emph{Additivity} Test Error(\%}) \\ \hline
$X_{1}$: \emph{IDQ\_MITE\_UOPS}  & 13 \\ \hline
$X_{2}$: \emph{IDQ\_MS\_UOPS} & 37 \\ \hline
$X_{3}$: \emph{ICACHE\_64B\_IFTAG\_MISS} & 36 \\ \hline
$X_{4}$: \emph{ARITH\_DIVIDER\_COUNT} & 80 \\ \hline
$X_{5}$: \emph{L2\_RQSTS\_MISS} & 14 \\ \hline
$X_{6}$: \emph{FP\_ARITH\_INST\_RETIRED\_DOUBLE} & 11 \\ \hline
\end{tabular}
&
\begin{tabular}{llllllll}
                            & $E_{D}$                        & $X_{1}$                        & $X_{2}$                        & $X_{3}$                        & $X_{4}$                        & $X_{5}$                        & $X_{6}$                        \\  
\multicolumn{1}{l}{$E_{D}$} & \multicolumn{1}{l|}{1}         & \multicolumn{1}{l|}{0.53} & \multicolumn{1}{l|}{0.50} & \multicolumn{1}{l|}{0.42}  & \multicolumn{1}{l|}{0.58} & \multicolumn{1}{l|}{0.99} & \multicolumn{1}{l}{0.99} \\ \cline{2-8} 
\multicolumn{1}{l}{$X_{1}$} & \multicolumn{1}{l|}{0.53} & \multicolumn{1}{l|}{1}         & \multicolumn{1}{l|}{0.41} & \multicolumn{1}{l|}{0.25} & \multicolumn{1}{l|}{0.39} & \multicolumn{1}{l|}{0.45} & \multicolumn{1}{l}{0.44} \\ \cline{2-8} 
\multicolumn{1}{l}{$X_{2}$} & \multicolumn{1}{l|}{0.50} & \multicolumn{1}{l|}{0.41} & \multicolumn{1}{l|}{1}         & \multicolumn{1}{l|}{0.19} & \multicolumn{1}{l|}{0.99} & \multicolumn{1}{l|}{0.48} & \multicolumn{1}{l}{0.48} \\ \cline{2-8} 
\multicolumn{1}{l}{$X_{3}$} & \multicolumn{1}{l|}{0.42}  & \multicolumn{1}{l|}{0.25} & \multicolumn{1}{l|}{0.19} & \multicolumn{1}{l|}{1}         & \multicolumn{1}{l|}{0.21} & \multicolumn{1}{l|}{0.41} & \multicolumn{1}{l}{0.40} \\ \cline{2-8} 
\multicolumn{1}{l}{$X_{4}$} & \multicolumn{1}{l|}{0.58} & \multicolumn{1}{l|}{0.39} & \multicolumn{1}{l|}{0.99} & \multicolumn{1}{l|}{0.21} & \multicolumn{1}{l|}{1}         & \multicolumn{1}{l|}{0.57} & \multicolumn{1}{l}{0.56} \\ \cline{2-8} 
\multicolumn{1}{l}{$X_{5}$} & \multicolumn{1}{l|}{0.99} & \multicolumn{1}{l|}{0.45} & \multicolumn{1}{l|}{0.48} & \multicolumn{1}{l|}{0.41} & \multicolumn{1}{l|}{0.57} & \multicolumn{1}{l|}{1}         & \multicolumn{1}{l}{0.99} \\ \cline{2-8} 
\multicolumn{1}{l}{$X_{6}$} & \multicolumn{1}{l|}{0.99} & \multicolumn{1}{l|}{0.44} & \multicolumn{1}{l|}{0.48} & \multicolumn{1}{l|}{0.40} & \multicolumn{1}{l|}{0.56} & \multicolumn{1}{l|}{0.99} & \multicolumn{1}{l}{1}         \\ \cline{2-8} 
\end{tabular} \\
& \\
\textbf{A} & \textbf{B}
\end{tabular}
\end{table*}

\begin{table*}[!t]
	\centering
	\caption{Linear predictive models ($A_1$-$G_1$) with intercepts and their minimum, average, and maximum prediction errors. Coefficients can be positive or negative.}
	\scriptsize
	\label{tab:models-with-intercepts}
	\begin{tabular}{|p{.5cm}|p{3cm}|p{8cm}|p{2.5cm}|}
		\hline
		\textbf{Model} & \textbf{PMCs} & \textbf{Intercept followed by Coefficients} & \textbf{Percentage prediction errors (min, avg, max)} \\ \hline
		$A_1$ & $X_1,X_2,X_3,X_4,X_5,X_6$ &  1.02E+01, 3.06E-09, 1.95E-08, 3.30E-07, -1.02E-06, 6.18E-08, -9.39E-11 & (2.7, 32, 99.9) \\ \hline
		$B_1$ & $X_1,X_2,X_3,X_5,X_6$ & 1.28E+01, 3.68E-09, 2.26E-10, 3.43E-07, 7.40E-08, -4.763E-10 & (2.5, 23.32, 80.42) \\ \hline
		$C_1$ & $X_1,X_3,X_5,X_6$ & 1.64E+01, 3.71E-09, 3.34E-07, 7.45E-08, -4.87E-10 & (2.5, 21.86, 76.9) \\ \hline
		$D_1$ & $X_1,X_5,X_6$ & 2.99E+01, 3.72E-09, 7.54E-08, -5.076E-10 & (2.5, 21.78, 77.33) \\ \hline
		$E_1$ & $X_1,X_6$ & 1.30E+02, 4.21E-09, 1.456E-09 & (2.5, 18.01, 89.23) \\ \hline
		$F_1$ & $X_6$ & 7.49E+02, 1.53E-09 & (2.5, 14.39, 34.64) \\ \hline
		$G_1$ & $X_4,X_5,X_6$ & 4.92E+02, 6.79E-08, 9.45E-08, -9.60E-10 & (2.5, 23.46, 80) \\ \hline
	\end{tabular}
\end{table*}

\begin{table*}[!t]
	\centering
	\caption{Linear predictive models ($A_2$-$G_2$) with zero intercepts and their minimum, average, and maximum prediction errors. Coefficients can be positive or negative.}
	\scriptsize
	\label{tab:models-with-zero-intercepts}
	\begin{tabular}{|p{.5cm}|p{3cm}|p{8cm}|p{2.5cm}|}
		\hline
		\textbf{Model} & \textbf{PMCs} & \textbf{Coefficients} & \textbf{Percentage prediction errors (min, avg, max)} \\ \hline
		$A_2$ & $X_1,X_2,X_3,X_4,X_5,X_6$ & 1.08E-09, 1.96E-08, 3.51E-07, -1.02E-06, 6.19E-08, -9.78E-11  & (2.5, 32, 78.7) \\ \hline
		$B_2$ & $X_1,X_2,X_3,X_5,X_6$ & 3.71E-09, 2.37E-10, 3.69E-07, 7.42E-08, -4.82E-10 & (2.5, 23.32, 80.57) \\ \hline
		$C_2$ & $X_1,X_3,X_5,X_6$ & 3.75E-09, 3.66E-07, 7.48E-08, -4.95E-10 & (2.5, 22.1, 77.5) \\ \hline
		$D_2$ & $X_1,X_5,X_6$ & 3.80E-09, 7.61E-08, -5.27E-10 & (2.5, 22.4, 78.5) \\ \hline
		$E_2$ & $X_1,X_6$ & 4.60E-09, 1.46E-09 & (2.5, 18.01, 89.45) \\ \hline
		$F_2$ & $X_6$ & 1.60E-09 & (3.0, 68.53, 90.53) \\ \hline
		$G_2$ & $X_4,X_5,X_6$ & 1.34E-07, 1.22E-07, -1.65E-09 & (2.5, 47.5, 111.22) \\ \hline
	\end{tabular}
\end{table*}

\begin{table*}[!t]
\centering
\caption{Linear predictive models ($A_3$-$G_3$) with zero intercepts. Coefficients cannot  be negative. The minimum, average, and maximum prediction errors of \emph{IntelRAPL} and the linear predictive models.}
\scriptsize
\label{tab:models-with-zero-intercepts-pos-coeffs}
\begin{tabular}{|p{1cm}|p{3cm}|p{7cm}|p{2.5cm}|}
\hline
\textbf{Model} & \textbf{PMCs} & \textbf{Coefficients} & \textbf{Percentage prediction errors (min, avg, max)} \\ \hline
$A_3$ & $X_1,X_2,X_3,X_4,X_5,X_6$ & 3.83E-09, 3.67E-10, 5.30E-07, 0.00E+00, 5.56E-08, 0.00E+00 & (6.6, 31.2, 61.9) \\ \hline
$B_3$ & $X_1,X_2,X_3,X_5,X_6$ & 3.83E-09, 3.67E-10, 5.30E-07, 0.00E+00, 5.56E-08 & (6.6, 31.2, 61.9) \\ \hline
$C_3$ & $X_1,X_3,X_5,X_6$ & 3.75E-09, 5.34E-07, 5.58E-08, 0.00E+00 & (2.5, 25.3, 62.1) \\ \hline
$D_3$ & $X_1,X_5,X_6$ & 4.00E-09, 5.59E-08, 0.00E+00 & (2.5, 23.86, 100.3) \\ \hline
$E_3$ & $X_1,X_6$ & 4.60E-09, 1.46E-09 & (2.5, 18.01, 89.45) \\ \hline
$F_3$ & $X_6$ & 1.60E-09 & (2.5, 68.5, 90.5) \\ \hline
$G_3$ & $X_4,X_5,X_6$ & 1.72E-07, 5.86E-08, 0.00E+00 & (2.5, 50, 77.9) \\ \hline
\emph{IntelRAPL} & & & (4.1, 30.6, 58.9) \\ \hline
\end{tabular}
\end{table*}

We build three types of linear regression models as follows:
\begin{itemize}
\item \textbf{Type 1}: Models $A_1$-$G_1$ with no restrictions on intercepts and coefficients.
\item \textbf{Type 2}: Models $A_2$-$G_2$ whose intercepts are forced to zero. 
\item \textbf{Type 3}: Models $A_3$-$G_3$ whose intercepts are forced to zero and whose coefficients cannot be negative.
\end{itemize}

Within each type $t$, $A_t$ employs all the PMCs as predictor variables. $B_t$ is based on five PMCs with the least additive PMC ($X_4$) removed. $C_t$ uses four PMCs with two most non-additive PMCs ($X_2,X_4$) removed and so on until $F_t$ containing only the most additive PMC ($X_6$). $G_t$ uses three PMCs ($X_4,X_5,X_6$) with the highest correlation with dynamic energy consumption.

For constructing all the models, we use a dataset of 277 points where each point  
contains dynamic energy consumption and the PMC counts for execution of one base application from Table \ref{table:applications} with some particular input. For testing the prediction accuracy of the models, we construct a test dataset of 50 different compound applications. We used this division (227 for training, 50 for testing) based on best practices and experts' opinion in this domain.

Table \ref{tab:models-with-intercepts} summarizes the type 1 models. Following are the salient observations:
\begin{itemize}
	\item The model intercepts are significant. In our theory of \emph{energy of computing} where we consider modelling of dynamic energy consumption, the intercepts are not present since they have no real physical meaning. Consider the case where no application is executed. The values of the PMCs will be zero and therefore the models must output the dynamic energy consumption to be zero. The models however output the values of their intercepts as the dynamic energy consumption. This violates the energy conservation property in the theory.
	\item $A_1$ has negative coefficients for PMCs, $X_4$ and $X_6$. Models $B_1$-$D_1$ have negative coefficients for PMC, $X_6$. The negative coefficients in these models can give rise to negative predictions for applications where the counts for $X_4$ and $X_6$ are higher than the other PMCs. We illustrate this case by designing a microbenchmark that stresses specifically hardware components resulting in large counts for the PMCs with the negative coefficients. Since, in our case, $X_4$ and $X_6$ count the division and floating point instructions, our microbenchmark is a simple \textit{assembly} language program that performs floating point division operations in a loop. When run for forty seconds, the PMC counts for this application on our platform were: $X_1$=7022011, $X_2$=623142, $X_3$=121489, $X_4$=5101219180, $X_5$=33210, and $X_6$=186971207082. The energy consumption predictions for this application from our four models $\{A_1, B_1, C_1, D_1\}$ are $\{-5210.52, -76.23, -74.59, -64.98\}$ which violate the energy conservation law.
	\item Since the predictor variables have a high positive correlation with energy consumption, their coefficients should exhibit the same relationship. The coefficients however have different signs for different models. Consider, for example, $X_{4}$ in $A_1$ and $C_1$. While it has positive coefficient in $A_1$, it has a negative coefficient in $C_1$. Similarly, $X_{6}$ in $A_1$ and $B_1$ has negative coefficient, whereas in $F_1$ it has a positive coefficient. We have found that the research works that propose linear models using these PMCs do not contain any sanity check for these coefficients. Therefore, we believe that using them in models without understanding the true meaning or the nature of their relationship with dynamic energy consumption can lead to serious inaccuracy.
\end{itemize}

The type 2 models are built using specialized linear regression, which forces the intercept to be zero. Table \ref{tab:models-with-zero-intercepts} contains their summary. All the models excepting $E_2$ and $F_2$ contain negative coefficients and therefore present the same issues that violate the energy conservation law.

The type 3 models are built using penalized linear regression using \textit{R programming} interface that forces the coefficients to be non-negative. All the models of this type have zero intercept and are summarized in the Table \ref{tab:models-with-zero-intercepts-pos-coeffs}. They incorporate basic sanity checks that disallow violations of energy conservation property.

We will now focus on the minimum, average, and maximum prediction errors of type 3 models. They are (6.6\%, 31.2\%, 61.9\%) for $A_3$. Since the coefficients are constrained to be non-negative, $X_6$ ends up having a zero coefficient. We remove the PMC with the next highest non-additivity ($X_4$) and construct $B_3$ based on the remaining five PMCs. In this model, $X_5$ has a zero coefficient. Its prediction errors are (6.6\%, 31.2\%, 61.9\%). We then remove the PMC with the next highest non-additivity ($X_2$) from the list of four and build $C_3$ based on the remaining PMCs. Its prediction errors are (2.5\%, 25.3\%, 62.1\%). Finally, we build $F_3$ with just one most additive PMC ($X_6$). Its prediction errors are (2.5\%, 68.5\%, 90.5\%). The prediction errors of RAPL are (4.1\%, 30.6\%, 58.9\%). The prediction errors of $G_3$ are (2.5\%, 50\%, 77.9\%).

We derive the following conclusions:
\begin{itemize}
	\item As we remove \emph{non-additive} PMCs one by one, the average prediction accuracy of the models improves significantly. $E_3$ with two most additive PMCs is the best in terms of average prediction accuracy. We therefore conclude that employing \emph{non-additive} PMCs can significantly impair the prediction accuracy of models and that inclusion of highly \emph{additive} PMCs improves the prediction accuracy of models drastically. 	
	\item We highlight two examples demonstrating the dangers of pure fitting exercise (for example: applying linear regression) without understanding the true physical significance of a parameter.
	\begin{itemize}
		\item The PMC $X_6$, which has the highest significance in terms of contribution to dynamic energy consumption (highest additivity), ends up having a zero coefficient in $A_3$, $C_3$, $D_3$, and $G_3$. $D_3$ has only two PMCs, $X_1$ and $X_5$, effectively. The linear fitting method picks $X_5$ instead of $X_6$ thereby impairing the prediction accuracy of $D_3$ (and also $G_3$). This is because $X_5$ and $X_6$ have high positive correlation between themselves but the fitting method does not know that $X_6$ is highly additive.
		\item $F_3$ containing one PMC with the highest additivity, $X_6$, has the lowest prediction accuracy. The linear fitting method is unable to find a good fit.
    \end{itemize}
	\item The average prediction accuracy of \emph{RAPL} is equal to that of the $A_3$ and $B_3$, which contain the highest number of \emph{non-additive} PMCs. If the model of \emph{RAPL} is disclosed, one can check how much its prediction accuracy can be improved by removing \emph{non-additive} PMCs and including highly \emph{additive} PMCs.
	\item $G_3$ fares worse than \emph{RAPL} and $A_3$ even though it contains PMCs that are highly correlated with dynamic energy consumption. $E_3$ with two most additive PMCs has better average prediction accuracy than $G_3$, which demonstrates that \emph{additivity} is a more important criterion than correlation. 
\end{itemize}

\begin{figure*}[!t]
\centering
\includegraphics[width=5in,height=2.7in]{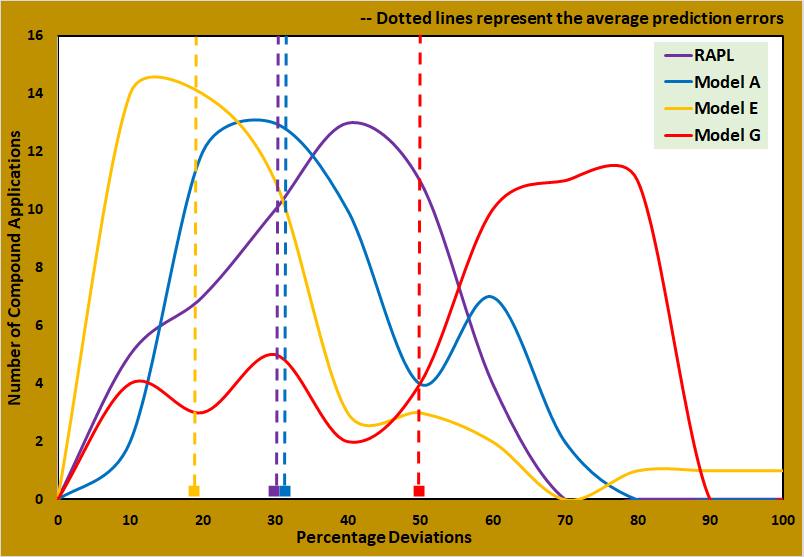}
\caption{Percentage deviations of the type 3 models shown in Table \ref{tab:models-with-zero-intercepts-pos-coeffs} from the system-level physical measurements provided by power meters (\emph{HCLWattsUp}). The dotted lines represent the averages.}
\label{fig:accuracytest}
\end{figure*}

Figure \ref{fig:accuracytest} presents the percentage deviations in dynamic energy consumption predictions by type 3 models (Table \ref{tab:models-with-zero-intercepts-pos-coeffs}) from the system-level physical measurements obtained using \emph{HCLWattsUp} (using WattsUp Pro power meters) for different compound applications. $RAPL$, $A_3$, and $G_3$ exhibit higher average percentage deviations than the best model, $E_3$. While $RAPL$ distribution is normal, $A_3$ and $G_3$ demonstrate non-normality suggesting systemic (not fully random) deviations from the average.

\subsection{Study of Dynamic Energy Optimization using \emph{IntelRAPL} and System-level Physical Measurements}

In this section, we demonstrate that using inaccurate energy measuring tools in energy optimization methods may lead to significant energy losses.

We study optimization of a parallel matrix-matrix multiplication application for dynamic energy using two measurement tools, \emph{IntelRAPL} \cite{IntelRAPL} which is a popular mainstream tool and system-level physical measurements using power meters (\emph{HCLWattsUp} \cite{HCLWattsUp}) which we believe are accurate. 

\begin{table}[!t]
	\footnotesize
	\caption{Specification of the Intel Skylake multicore CPU}
	\label{table:intel-skylake-multicore-CPU}
	\centering
	\begin{tabular}{|l|l|}
		\hline 
		\textbf{Technical Specifications} & \textbf{Intel Skylake Server} \\ \hline
		Processor & \textbf{Intel(R) Xeon(R) Gold 6152} \\ \hline
		OS & Ubuntu 16.04 LTS \\ \hline
		Micro-architecture & Skylake \\ \hline
		Thread(s) per core & 2 \\ \hline
		Socket(s) & 1 \\ \hline
		Cores per socket & 22 \\ \hline
		NUMA node(s) & 1 \\ \hline
		L1d cache &  32 KB \\ \hline
		L11 cache &  32 KB \\ \hline
		L2 cache &  1024 KB \\ \hline
		L3 cache &  30976 KB \\ \hline
		Main memory &  96 GB \\ \hline
		TDP & 140 W \\ \hline	
		Idle Power & 32 W \\ \hline
	\end{tabular}
\end{table}

For this purpose, we employ a data-parallel application that uses \emph{Intel MKL DGEMM} as building block. The experimental platform consists of two servers, HCLserver1 (Table \ref{table:intel-haswell-multicore-CPU}) and HCLserver2 (Table \ref{table:intel-skylake-multicore-CPU}). To find the partitioning of matrices between the servers that minimizes the dynamic energy consumption, we use a model-based data partitioning algorithm, which takes as input dynamic energy functional models of the servers. We compare the total dynamic energy consumptions of the solutions returned when the input dynamic energy models of the servers are built using \emph{IntelRAPL} \cite{IntelRAPL} and 
HCLWattsUp \cite{HCLWattsUp}.
We follow the same strict experimental methodology as in the previous experimental setup to make sure that our experimental results are reliable.

The parallel application computes a matrix product of two dense square matrices $A$ and $B$ of sizes $N \times N$ and is executed using two processors, \emph{HCLserver1} and \emph{HCLserver2}. The matrix $A$ is partitioned between the processors as $A_1$ and $A_2$ of sizes $M \times N$ and $K \times N$  where $M+K=N$. Matrix $B$ is replicated at both the processors. Processor \emph{HCLserver1} computes the product of matrices $A_1$ and $B$ and processor \emph{HCLserver2} computes the product of matrices $A_2$ and $B$. There are no communications involved.

\begin{figure*}
	\centering
	\begin{tabular}{@{}c@{}}
		\includegraphics[width=0.9\linewidth]{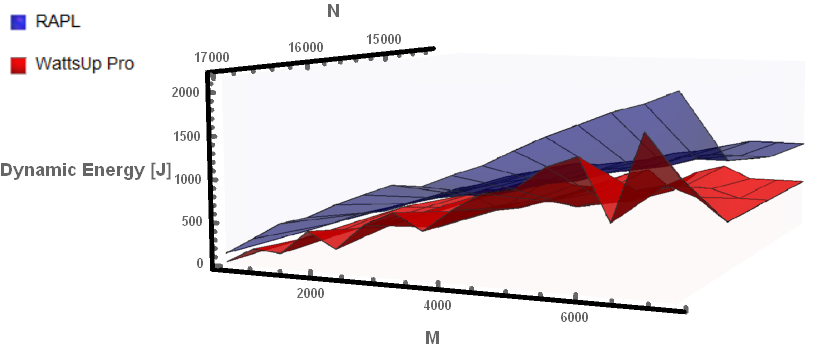} \\[\abovecaptionskip]
		\small \textbf{(a)} 
	\end{tabular}
	
	\begin{tabular}{@{}c@{}}
		\includegraphics[width=0.9\linewidth]{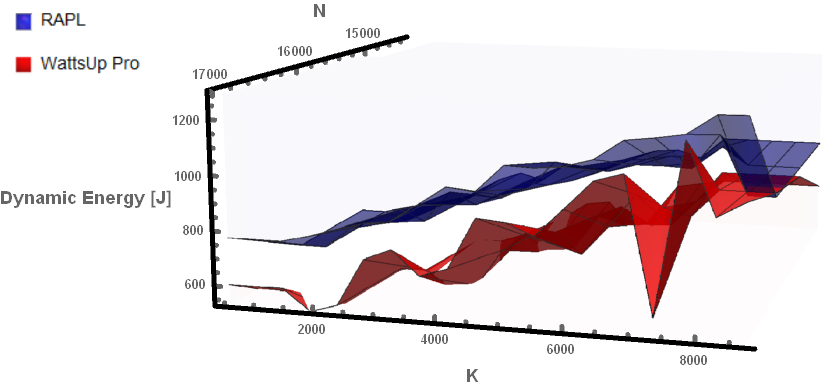} \\[\abovecaptionskip]
		\small \textbf{(b)} 
	\end{tabular}
	
	\caption{Dynamic energy consumption of Intel MKL DGEMM application multiplying two matrices of sizes \textbf{(a)} $M \times N$ and $N \times N$ on HCLServer1, and \textbf{(b)} $K \times N$ and $N \times N$ on HCLServer2. $M+K=N$.}
	\label{fig:dgemmserver}	
\end{figure*}

The decomposition of the matrix $A$ is computed using a model-based data partitioning algorithm. The inputs to the algorithm are the number of rows of the matrix $A$, $N$, and the dynamic energy consumption functions of the processors, $\{E_1,E_2\}$. The output is the partitioning of the rows, $(M,K)$. The discrete dynamic energy consumption function of processor $P_i$ is given by $E_i=\{e_i(x_1,y_1),...,e_i(x_m,y_m)\}$ where $e_i(x,y)$ represents the dynamic energy consumption during the matrix multiplication of two matrices of sizes $x \times y$ and $y \times y$ by the processor $i$. Figure \ref{fig:dgemmserver} shows the discrete dynamic energy consumption functions of \emph{IntelRAPL} and \emph{HCLWattsUp} for the processors \emph{HCLserver1} and \emph{HCLserver2}. The dimension $y$ ranges from 14336 to 16384 in steps of 512. For \emph{HCLserver1}, the dimension $x$ ranges from 512 to $y/2$ in increments of 512. For \emph{HCLserver2}, the dimension $x$ ranges from $y-512$ to $y/2$ in decrements of 512.

The main steps of the data partitioning algorithm are as follows:

\textbf{1. \textbf{Plane intersection of dynamic energy functions}:} Dynamic energy consumption functions $\{E_1,E_2\}$ are cut by the plane $y=N$ producing two curves that represent the dynamic energy consumption functions against $x$ given $y$ is equal to $N$.

\textbf{2. \textbf{Determine $M$ and $K$}:} 

\begin{equation*} 
(M,K) = \argmin_{\substack{M \in (512,N/2), \\ K \in (N-512,N/2), \\ M+K=N}} (e_1(M,N) + e_2(K,N)) 
\end{equation*}

We use four workload sizes $\{14336,14848,15360,16384\}$ in our test data. For each workload size, we determine the workload distribution using the data partitioning algorithm employing model based on \emph{IntelRAPL}. We execute the parallel application using this workload distribution and determine its dynamic energy consumption. We represent it as $e_{rapl}$. We obtain the workload distribution using the data partitioning algorithm employing model based on \emph{HCLWattsUp}. We execute the parallel application using this workload distribution and determine its dynamic energy consumption. We represent it as $e_{hclwattsup}$. We calculate the percentage loss of dynamic energy consumption provided by \emph{HCLWattsUp} compared to \emph{IntelRAPL} as $(e_{rapl} - e_{hclwattsup})/e_{hclwattsup} \times 100$. Losses for the four workload sizes are $\{65,58,56,56\}$.

\section{Related Work}

This section presents a brief literature survey of some important tools widely used to obtain PMCs, notable research on energy predictive models, and research works that provide a critical review of PMCs.

\textit{Tools to obtain PMCs}. \emph{Perf} \cite{LinuxPerf} can be used to gather the PMCs for CPUs in Linux. PAPI \cite{PAPI} and Likwid \cite{treibig2010likwid} allow obtaining PMCs for Intel and AMD microprocessors. \emph{Intel PCM} \cite{IntelPCM} gives PMCs of core and uncore components of an Intel processor. For Nvidia GPUs, CUDA Profiling Tools Interface (\emph{CUPTI}) \cite{CUPTI} can be used for obtaining the PMCs.

\textit{Notable Energy Predictive Models for CPUs}. Initial Models correlating PMCs to energy values include \cite{Bellosa2000, IcsiMartonosi2003, Li2003, Lee2006, Heath2005, Economou2006, Fan2007, Kansal2008}. Events such as integer operations, floating-point operations, memory requests due to cache misses, component access rates, instructions per cycle (IPC), CPU/disk and network utilization,  etc. were believed to be strongly correlated with energy consumption. Simple linear models have been developed using PMCs and correlated features to predict energy consumption of platforms. Rivoire et al. \cite{Rivoire2008A, Rivoire2008B} study and compare five full-system real-time power models using a variety of machines and benchmarks. They report that PMC-based model is the best overall in terms of accuracy since it accounted for majority of the contributors to system's dynamic power. Other notable PMC-based linear models are \cite{Singh2009, Powell2009, Goel2010, Wang2010, Basmadjian2011, Dargie2015, haj2016fine }  

Rotem et al. \cite{IntelRAPL} present \emph{RAPL}, in Intel Sandybridge to predict the energy consumption of core and uncore components (QPI, LLC) based on some PMCs (which are not disclosed). Lastovetsky et al. \cite{Reddy2016TPDS} present an application-level energy model where the dynamic energy consumption of a processor is represented by a function of problem size.

\textit{Critiques of PMCs for Energy Predictive Modelling}. Some attempts where poor prediction accuracy of PMCs for energy predictive modeling has been critically examined include \cite{Economou2006, McCullough2011, hackenberg2013power, Ken2017}. Researchers highlight the fundamental limitation to obtain all the PMCs simultaneously or in one application run and show that linear regression models give prediction errors as high as 150\%. The property of additivity of PMCs is first introduced in \cite{shahid2017additivity}.

\section{Conclusion}

Energy predictive modelling based on PMCs is now the leading method for prediction of energy consumption during an application execution. We summarized the assumptions behind the existing models and used a model-theoretic approach to formulate their assumed properties in a mathematical form. We extended the formalism by adding properties, heretofore unconsidered, that are basic implications of the universal energy conservation law. The extended formalism forms our theory of \emph{energy of computing}.

We considered practical implications of our theory and applied them to improve the prediction accuracy of the state-of-the-art energy predictive models. First implication concerns studying \emph{additivity} of model parameters. We studied the \emph{additivity} of PMCs on a modern Intel platform. We showed that a PMC can be \emph{non-additive} with error as high as 3075\% and there are PMCs where the error is over 100\%.

We selected six PMCs which are common in the state-of-the-art energy predictive models and which are highly correlated with dynamic energy consumption. We constructed seven linear regression models with the PMCs as predictor variables and that pass the constraints. We demonstrated that prediction accuracy of the models improves as we remove one by one from them highly \emph{non-additive} PMCs. We also highlighted the drawbacks of pure fitting exercise (for example: applying linear regression) without understanding the true physical significance of a parameter. We show that linear regression methods select PMCs based on high positive correlation with dynamic energy consumption and ignore PMCs that have high significance in terms of contribution to dynamic energy consumption (due to high additivity) thereby impairing the prediction accuracy of the models.

Finally, we studied optimization of a parallel matrix-matrix multiplication application for dynamic energy using two measurement tools, \emph{IntelRAPL} \cite{IntelRAPL}, which is a popular mainstream tool, and power meters (\emph{HCLWattsUp} \cite{HCLWattsUp}) providing accurate system-level physical measurements. We demonstrated that we lose significant amount of energy (up to 67\% for applications used in the experiments) by using \emph{IntelRAPL} most likely because it does not take into account the \emph{energy conservation} properties (we found no explicit evidence that it does).

\section*{Appendix}

\textbf{Appendix A: Rationale Behind Using Dynamic Energy Consumption Instead of Total Energy Consumption} \label{denergy-rationale}

We consider only the dynamic energy consumption in our work for reasons below:

\begin{enumerate}
	\item Static energy consumption is a constant (or a inherent property) of a platform that can not be optimized. It does not depend on the application configuration.
	\item Although static energy consumption is a major concern in embedded systems, it is becoming less compared to the dynamic energy consumption due to advancements in hardware architecture design in HPC systems.
	\item We target applications and platforms where dynamic energy consumption is the dominating energy dissipator.
	\item Finally, we believe its inclusion can underestimate the true worth of an optimization technique that minimizes the dynamic energy consumption. We elucidate using two examples from published results.
	\begin{itemize}
		\item In our first example, consider a model that reports predicted and measured total energy consumption of a system to be 16500J and 18000J. It would report the prediction error to be 8.3\%. If it is known that the static energy consumption of the system is 9000J, then the actual prediction error (based on dynamic energy consumptions only) would be 16.6\% instead.
		\item In our second example, consider two different energy prediction models ($M_A$ and $M_B$) with same prediction errors of 5\% for an application execution on two different machines ($A$ and $B$) with same total energy consumption of 10000J. One would consider both the models to be equally accurate. But supposing it is known that the dynamic energy proportions for the machines are 30\% and 60\%. Now, the true prediction errors (using dynamic energy consumptions only) for the models would be 16.6\% and 8.3\%. Therefore, the second model $M_B$ should be considered more accurate than the first.
	\end{itemize}
\end{enumerate}

\ifCLASSOPTIONcompsoc
\section*{Acknowledgments}
\else
\section*{Acknowledgment}
\fi

This publication has emanated from research conducted with the financial support of Science Foundation Ireland (SFI) under Grant Number 14/IA/2474.

\bibliographystyle{IEEEtran}
\bibliography{IEEEabrv,paper}

\end{document}